\documentclass[%
reprint,
 showpacs,
 amsmath,amssymb,
]{revtex4-1}
\usepackage{romannum}
\usepackage{graphicx}
\usepackage{dcolumn}
\usepackage{bm}

\usepackage{CJK}

\usepackage{amsmath}
\usepackage{bm}  

\usepackage{color}
\usepackage{amssymb}
\usepackage{amsfonts}

\definecolor{Red}{rgb}{1,0,0}

\usepackage[colorlinks,urlcolor=blue,linkcolor=blue,citecolor=blue,anchorcolor=blue]{hyperref}
\makeatletter

\newcommand{\Rmnum}[1]{\expandafter\@slowromancap\romannumeral #1@}
\makeatother
\begin{document}

\preprint{APS/123-QED}

\title{Generation of energy-time entangled triphotons in a six-level cold atomic system}


\author{Ling Niu$^{1}$}
\author{Zhiyin Duan$^{1}$}
\author{Na Liu$^{1}$}
\author{Yitong Zhai$^{1}$}
\author{Shaoyan Liu$^{1}$}
\author{Junsheng Li$^{3}$}
\email{lijs@sxnu.edu.cn}
\author{Donghai Zhang$^{3}$}
\email{zhangdh@sxnu.edu.cn}
\author{Da Zhang$^{1,2}$}
\email{zhang1556433@sxnu.edu.cn}

\affiliation{%
$^1$\mbox{School of Physics and Information Engineering, Shanxi Normal University, Taiyuan 030031, China} \\
$^2$\mbox{Key Laboratory of Magnetic Molecules and Magnetic Information Materials, Ministry of Education, Taiyuan 030031, China} \\
$^3$\mbox{Institute of Modern Physics, Shanxi Normal University, Taiyuan 030031, China} \\
}%
\date{\today}

\begin{abstract}
\noindent
Multiphoton entangled states are pivotal resources for implementing optical quantum information protocols.
Recently, energy-time-entangled triphotons have been observed in hot atomic ensembles.
However, in these protocols, the complex fifth-order nonlinear susceptibility entailed by four- or five-level systems limits our understanding of triphoton generation.
Here, to directly capture the generation mechanism of triphotons and their associated optical properties, we investigate the generation of energy-time-entangled triphotons in a six-level cold atomic ensemble.
The fifth-order nonlinear susceptibility indicates the existence of two sets of spontaneous six-wave mixing in the system.
Notably, triphoton generation in this system is subject to stringent timing constraints.
Collectively, these characteristics give rise to threefold coincidence counts, which--dominated by the fifth-order nonlinear susceptibility--exhibit asymmetrically damped Rabi oscillations in the two-dimensional time domain.
Furthermore, we analytically derive that the temporal correlation properties of conditional two-photon states are preserved--a unique feature of $W$-class tripartite entanglement.
These results not only lay the groundwork for the experimental preparation of triphotons using six-level systems but also provide key support for understanding the generation mechanism of triphotons involving more complex fifth-order nonlinear susceptibilities.

\end{abstract}

\maketitle
\section{Introduction}
Quantum entanglement, one of the most iconic nonclassical features of quantum mechanics, transcends the conceptual framework of locality and realism in classical physics \cite{einstein.pr.47.777.1935}, standing as a core resource that drives the advancement of quantum information technology \cite{Pittman.pra.52.R3429.1995,Strekalov.prl.74.3600.1995,duan.nat.414.106500.2001,weedbrook.rmp.84.621.2012}.
Its unique nonlocal properties underpin progress across diverse fields.
In quantum communication, entangled states act as a fundamental resource underpinning secure communication protocols \cite{sheng.pra.82.032318.2010,gisin2007quantum}.
In quantum computing, multipartite entangled states form the cornerstone for implementing quantum parallel computing and achieving exponential speedups \cite{walther.nat.434.169.2005,deng.prl.130.190601.2023}.
In quantum precision metrology, entanglement-enhanced sensitivity enables surpassing the standard quantum limit \cite{Vittorio.PRL.96.010401.2006,lu.prl.129.070502.2022,Joo.prl.107.083601.2011}, thereby facilitating high-precision detection of weak physical quantities, with accuracy unattainable by classical measurement techniques.

Owing to their low decoherence rate, photons function as robust entanglement carriers in many experiments.
In general, the preparation of multiphoton entangled states primarily hinges on biphoton sources generated via spontaneous parametric down-conversion.
For instance, by using multiplexed biphoton sources and post-selection, polarization-entangled multiphoton states have been successfully prepared in succession \cite{bouwmeester1999observation,pan2001experimental}.
The energy-time and polarization-entangled three-photon states generated by cascade spontaneous parametric down-conversion have also been reported subsequently \cite{shalm.np.9.1.2012,hamel2014direct}.
Additionally, the preparation of multiphoton entangled sources encompasses the coherent mixing of entangled photon pairs with single photons attenuated from continuous-wave lasers \cite{mikami2005new,zhao.nature.430.54.2004}.
However, the bandwidth of multiphoton entanglement sources based on these techniques is on the order of GHz, rendering it challenging to interact effectively with atomic repeaters, whose linewidths are on the order of MHz.

Compared to crystal-based nonlinear optical systems, atomic ensembles exhibit unique advantages in preparing multiphoton entangled states.
Atoms exhibit well-defined energy level structures, strong nonlinearity, long coherence times, and multiple controllable parameters, enabling correlated photon pairs generated via spontaneous four-wave mixing to display long coherence times and high spectral brightness \cite{Harris.PRL.94.183601.2005,Harris.PRL.94.183601.2005,Harris.PRL.97.113602.2006,2016Subnatural}.
Notably, the linewidth of photon pairs generated from this model naturally matches that of quantum repeaters based on atomic ensembles, and such photon pairs have currently been applied to single-photon storage with 85\% efficiency \cite{Yunfei2019Efficient}, a single-photon precursor \cite{du.prl.106.243602.2011}, anti-parity-time symmetry \cite{PhysRevLett.123.193604}, and quantum key distribution \cite{Liuchang.ol.8.2013}.
Beyond these applications, the generation process of energy-time entangled photon pairs encompasses rich physical mechanisms.
The biphoton waveform is governed by the third-order nonlinear susceptibility and the phase mismatch function \cite{wen.PRA.77.033816.2008,wen.PRA.75.033809.2007}.
The former not only characterizes the strength of the nonlinear process but also reveals the existence of two sets of spontaneous four-wave mixing processes in the system, owing to its inherent double resonance structure \cite{wen.pra.74.023808.2006,wen.pra.74.023809.2006,wen.PRA.76.013825.2007}.
Destructive interference between these two sets of four-wave mixing gives rise to damped Rabi oscillations in two-photon coincidence counts.
Additionally, the linear optical response of correlated photon pairs can be dynamically tuned via the electromagnetically induced transparency effect, such that their amplitude is determined by the phase mismatch function \cite{du.prl.100.183603.2008,du.optica.2.2014}.
Atomic ensembles also exhibit a spontaneous six-wave mixing (SSWM) process: specifically, when three classical light beams propagate through the atomic medium, triplets $E_{s1}$, $E_{s2}$, and $E_{s3}$ are generated via the fifth-order nonlinear susceptibility \cite{li2024direct}.
This constitutes a technique for generating energy-time entangled triphotons in a single step, which has been recently observed in hot atomic ensembles \cite{li2024direct,feng2025observation}.
Although the generation of energy-time entangled triplets in systems with distinct energy levels has been studied via the perturbation chain model in \cite{kangkang.aqt.35.2020,wu.pra.112.013706.2025}, the complex resonance structure implicit in fifth-order nonlinear susceptibility renders analytical solution of threefold coincidence counts challenging.
This, in turn, results in insufficient profound understanding of its optical properties.
More importantly, we recently found that triphoton generation in a five-level system follows a strict temporal sequence \cite{zhang2025temporal}.
We thus question whether this phenomenon occurs only for this specific level system, or may also arise in other triphoton generation schemes.

In this work, we investigate the generation mechanism of energy-time entangled triphotons in a six-level cold atomic system.
Three primary considerations underpin the adoption of the six-level configuration.
First, the model for calculating the optical response of the atomic system is notably tractable: at most one optical transition exists between any two energy levels, and the accuracy of the calculated results can be cross-validated against those derived from the probability-amplitude model \cite{du.josab.25.12.2008,zhang.pra.96.053849.2017}.
Second, our six-level system facilitates a clear elucidation of the triphoton generation mechanism: the fifth-order nonlinear susceptibility reveals the presence of two sets of SSWM within the system.
Third, the centrosymmetric structure of the fifth-order nonlinear susceptibility gives rise to asymmetric damped Rabi oscillations in the threefold coincidence counts dominated by the fifth-order nonlinear response, a behavior that admits a clear analytical solution and indicates that triphoton generation follows a strict temporal order, consistent with our previous findings in the five-level atomic system \cite{zhang2025temporal}.
We further characterize the temporal correlation properties among the subsystems.
Two-photon coincidence counts between $E_{s1}$ and $E_{s2}$ exhibit damped Rabi oscillations, analogous to those of correlated photon pairs generated in three- or four-level systems \cite{wen.pra.74.023808.2006,wen.pra.74.023809.2006,wen.PRA.76.013825.2007}.
In contrast, $E_{s1}-E_{s3}$ coincidence counts exhibit a single-peak exponential decay profile, whereas $E_{s2}-E_{s3}$ coincidence counts display a standard exponential decay.
These results further confirm that the generated triphotons correspond to an energy-time entangled $W$-class state--an entangled state category that shares identical entanglement characteristics with the standard
$W$-state but differs from it in state function form \cite{PhysRevA.62.062314}.
These findings not only deepen the fundamental understanding of triphoton entanglement mechanisms, but also provide a theoretical and experimental foundation for the development of $W$-state-based quantum information technologies.

The structure of the paper is as follows.
In section \Romannum{2}, we present the expressions for triphoton states and threefold coincidence counts.
Section \Romannum{3} focuses on deriving the linear and nonlinear susceptibilities of the generated signal by solving the Heisenberg operator evolution equations governing the atomic operators, alongside an analysis of the mechanisms underlying its generation and propagation.
In section \Romannum{4}, we analytically derive the threefold coincidence counts and the conditional two-photon coincidence counts, which are dominated by the fifth-order nonlinear susceptibility.
Finally, section \Romannum{5} presents the conclusions.

\section{Three-photon state function}
\begin{figure}[t]
\centering
  \includegraphics[width=8cm]{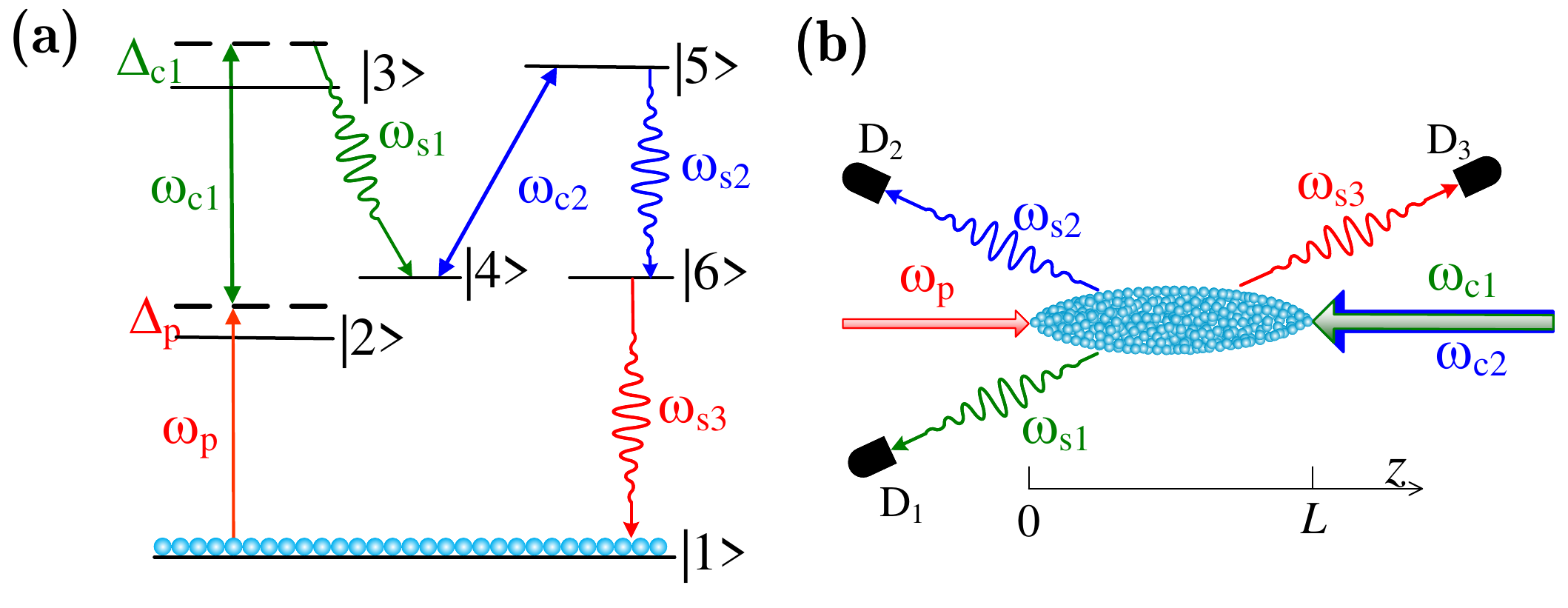}  
  \caption{(a) Generation of triplet photons in a six-level cold atom system. In the presence of the pump field and two coupling fields, triplet photons are spontaneously generated through the SSWM process. (b) Spatial configuration diagram of the incident light and generated signals.}
  \label{fig1}
\end{figure}

To begin our analysis, we focus on the generation of triple photons via SSWM in a six-level cold atomic system, as illustrated in Fig. \ref{fig1}(a).
These uniformly cooled atoms, confined within a thin cylindrical volume of length $L$ with atomic density $N$, are initially prepared in the ground state $|1\rangle$ through optical pumping.
Figure \ref{fig1}(b) depicts the spatial configuration of the classical and quantized fields involved in SSWM, in which a pump laser $E_p$ and two coupling lasers $E_{c1}$ and $E_{c2}$ propagate in counterpropagating directions through the medium.
A weak pump field $\omega_p$ drives the transition $|1\rangle\rightarrow|2\rangle$ with frequency detuning $\Delta_p=\omega_{21}-\omega_p$, a strong coupling field $\omega_{c1}$ induces the transition $|2\rangle\rightarrow|3\rangle$ with detuning $\Delta_{c1}=\omega_{32}-\omega_{c1}$, and a resonant strong coupling field $\omega_{c2}$ connects the atomic transition $|4\rangle\rightarrow|5\rangle$, where $\omega_{ij}$ is the transition frequency from level $|j\rangle$ to $|i\rangle$.
Driven by the fifth-order nonlinear susceptibility, phase-matched triple photons are generated via SSWM, adhering strictly to both energy and momentum conservation.

In the interaction picture, the SSWM is described by the effective Hamiltonian
\begin{eqnarray}\label{eq1}
\hat{H}_I=\varepsilon_0\int_Vd^3r\chi^{(5)}E_pE_{c1}E_{c2}\hat{E}^{(-)}_{s1}\hat{E}^{(-)}_{s2}\hat{E}^{(-)}_{s3}+H.c, \nonumber \\
\end{eqnarray}
where $\varepsilon_0$ denotes the permittivity of free space and \emph{V} is the interaction volume illuminated by the pump and two coupling beams simultaneously.
$\chi^{(5)}$ is the fifth-order nonlinear susceptibility defined by the nonlinear polarizability.
To simplify the calculations, the pump and two coupling beams are taken as the classical plane waves
\begin{eqnarray}\label{eq2}
E_p=E_{p0}e^{i(k_pz-\omega_pt)},~~E_{cj}=E_{cj0}e^{-i(k_{cj}z+\omega_{cj}t)}.
\end{eqnarray}
The $z$ direction is assumed to be parallel to the pump longitudinal propagation.
The generated $E_{sj}$ photons are given by the quantized fields
\begin{eqnarray}\label{eq3}
\hat{E}^{(+)}_{s3} &=& \sum_{\bm{k}_{s3}}E^{(+)}_{s3}\hat{a}_{\bm{k}_{s3}}e^{i(\bm{k}_{s3}\cdot\bm{r}-\omega_{s3}t)},    \nonumber \\
\hat{E}^{(+)}_{sj}& =& \sum_{\bm{k}_{sj}}E^{(+)}_{sj}\hat{a}_{\bm{k}_{sj}}e^{-i(\bm{k}_{sj}\cdot\bm{r}+\omega_{sj}t)}, (j=1,2)
\end{eqnarray}
where $E^{(+)}_{sj}=i\sqrt{\hbar\varpi_{sj}/(2\varepsilon_0n^2_{sj}V_q)}$ with the refraction index $n_{sj}$ and the quantization volume $V_q$.
$\varpi_{sj}$ denotes the central frequency of the $E_{sj}$ photon, whose real frequency is given by $\omega_{sj}=\varpi_{sj}+\delta_j$, where $\delta_j$ represents the small frequency shift of the $E_{sj}$ photon.
In our energy-level configuration, the three central frequencies corresponding to the signal fields are $\varpi_{s1}=\omega_{43}$, $\varpi_{s2}=\omega_{56}$, and $\varpi_{s3}=\omega_{61}$.
The wave vectors $\mathbf{k}_{si}$ are evaluated inside the material and $\hat{a}_{\mathbf{k}_{si}}$ are the annihilation operators at the output surface associated with $\mathbf{k}_{si}$.

Using Eqs. (\ref{eq2}) and (\ref{eq3}) and after integration,  the Hamiltonian (\ref{eq1}) can be rewritten as:
\begin{align}\label{eq4}
\hat{H}_I&=i\hbar\sum_{k_{s1}}\sum_{k_{s2}}\sum_{k_{s3}}\kappa\Phi(\Delta kL)H(\bm{\alpha}_{s1}+\bm{\alpha}_{s2}+\bm{\alpha}_{s3},\bm{\rho})  \nonumber \\
&\times\hat{a}_{k_{s1}}^\dag\hat{a}^\dag_{k_{s2}}\hat{a}^\dag_{k_{s3}}e^{-i(\omega_p+\omega_{c1}+\omega_{c2}-\omega_{s1}-\omega_{s2}-\omega_{s3})t}
\end{align}
where $\kappa=\sqrt{\hbar\varpi_{s1}\varpi_{s2}\varpi_{s3}/(8\varepsilon_0L)}\chi^{(5)}E_{p0}E_{c10}E_{c20}$ is the nonlinear parametric coupling coefficient.
$\Phi(\Delta kL)=(e^{i\Delta kL}-1)/(i\Delta kL)$ is the longitudinal detuning function, where $\Delta k=k_p-k_{c1}-k_{c2}+k_{s1}+k_{s2}-k_{s3}$ is the phase mismatch along the longitudinal axis.
Note that $\Phi(\Delta kL)$ carries the information of phase mismatch in the longitudinal direction over the entire medium.
$H(\bm{\alpha}_{s1}+\bm{\alpha}_{s2}+\bm{\alpha}_{s3},\bm{\rho})$ is called the transverse detuning function, which is the integral over the area $A$ of intersection of
the beam cross section
\begin{eqnarray}
H(\bm{\alpha}_{s1}+\bm{\alpha}_{s2}+\bm{\alpha}_{s3},\bm{\rho})=A^{-\frac{3}{2}}\int_{A}d^2\rho e^{-i(\bm{\alpha}_{s1}+\bm{\alpha}_{s2}+\bm{\alpha}_{s3})\cdot\bm{\rho}}.   \nonumber \\
\end{eqnarray}
In the derivation we assume that $A$ is independent of $z$.
$\bm{\alpha}_{si}$ is the transverse wave vector of the generated photon $\omega_{si}$ and $\bm{\rho}$ is in the transverse plane perpendicular to the longitudinal axis $z$.
Since we are interested in the temporal correlation of triple photons and the transverse detuning function has important applications in quantum imaging \cite{Alexander.prl.87.123602.2001}, we assume that the transverse area of the three input classical fields is large enough so that $H(\bm{\alpha}_{s1}+\bm{\alpha}_{s2}+\bm{\alpha}_{s3},\bm{\rho})$ can be approximated by a $\delta$ function.
The detailed derivation is presented in the Appendix.

Due to the weak nonlinear interaction, the triphoton state output at the cell surface, derived within the framework of first-order perturbation theory, is expressed as:
\begin{eqnarray}\label{eq5}
&&|\Psi\rangle =\frac{-i}{\hbar}\int_{-\infty}^{+\infty}dt \hat{H}_I|0\rangle \nonumber \\
&&=\sum_{k_{s1}}\sum_{k_{s2}}\sum_{k_{s3}}\kappa\Phi(\Delta kL)\delta(\mathfrak{\Delta}) \hat{a}_{k_{s1}}^\dag\hat{a}^\dag_{k_{s2}}\hat{a}^\dag_{k_{s3}}|0\rangle,
\end{eqnarray}
where $\mathfrak{\Delta}=\omega_p+\omega_{c1}+\omega_{c2}-\omega_{s1}-\omega_{s2}-\omega_{s3}$ and the Dirac $\delta(\mathfrak{\Delta})$ function comes from the time integral and states the energy conservation.
The state (\ref{eq5}) shows that the triphotons are entangled in both frequency and wave number.
Specifically, the full quantum state cannot be factorized into a product of any single-photon wave function and a two-photon wave function with respect to the wave numbers $k_1$, $k_2$, $k_3$.
For instance, it cannot be written in the form $f(k_i)f(k_j,k_l)$ for any permutation of $i,j,l$, which rigorously implies that the triphotons are entangled in the wave-number degree of freedom.
$\kappa$ is similar as shown the generation mechanism.

Our focus here is on the optical properties of the triplet photons, so we consider a simplified threefold coincidence counting experiment as illustrated in Fig. \ref{fig1}(b).
Detector $D_i$ register photons at frequencies $\omega_i$ at time $t_i$, respectively.
In particular, $D_1$ serves as a trigger, while $D_2$ and $D_3$ serve as stops.
Without loss of generality, the three detectors are positioned equidistant from the source, ensuring identical relative propagation time delays for the three output signals in free space.
With perfect detection efficiency, the average threefold coincidence counts rate is determined by \cite{wen.PRA.77.033816.2008,wen.74.023809.2006,wu.pra.112.013706.2025}
\begin{eqnarray}\label{eq6}
R_{cc}=\lim_{T\rightarrow\infty}\frac{1}{T}\int_{0}^{T}dt_1\int_{0}^{T}dt_2\int_{0}^{T}dt_{3}G^{(3)}
\end{eqnarray}
where $G^{(3)}=\langle\Psi|\hat{E}^{(-)}_{s1}\hat{E}^{(-)}_{s2}\hat{E}^{(-)}_{s3}\hat{E}^{(+)}_{s3}\hat{E}^{(+)}_{s2}\hat{E}^{(+)}_{s1}|\Psi\rangle$.
Since the spectral linewidth of the triplet photons generated by the atomic ensemble is on the order of MHz, which is much narrower than the GHz-scale spectral response range of single-photon detectors, the photon wavepacket remains nearly constant within the detector's response window.
Consequently, the detector can effectively measure the instantaneous intensity without requiring temporal averaging of the photon wavepacket.
Thus, Eq. (\ref{eq6}) can be simplified to
\begin{eqnarray}\label{eq7}
R_{cc}&&=|\langle0|\hat{E}^{(+)}_{s3}\hat{E}^{(+)}_{s2}\hat{E}^{(+)}_{s1}|\Psi\rangle|^2=|\mathfrak{B}(\tau_{12},\tau_{13})|^2,
\end{eqnarray}
where $\tau_{12}=t_2-t_1$ and $\tau_{13}=t_3-t_1$.
The function $\mathfrak{B}(\tau_{12},\tau_{13})$ is referred to as the triphoton waveform.
Substituting Eq. (\ref{eq5}) into Eq. (\ref{eq7}), one obtains
\begin{align}\label{eq8}
\mathfrak{B}(\tau_{12},\tau_{13})=B_2\int\int d\delta_2d\delta_3\chi^{(5)}\Phi(\Delta kL)e^{-i(\delta_2\tau_{12}+\delta_3\tau_{13})},
\end{align}
where all slowly varying terms are absorbed into $B_2$.
In Eq. (\ref{eq8}), we use $\sum_{k_j}\rightarrow V_q^{1/3}/(2\pi)\int d\omega_j/\nu_j$ to convert the summation into an integral, where $\nu_j$ represents the group velocity of the $E_{sj}$ photon inside the medium.

Equation (\ref{eq8}) reveals that the triphoton waveform is the two-dimensional Fourier transform of the product of the $\chi^{(5)}$ and $\Phi(\Delta kL)$.
This means that both $\chi^{(5)}$ and $\Phi(\Delta kL)$ are functions of $\delta_2$ and $\delta_3$.
Let us first consider the two-photon scenario.
The biphoton waveform of the correlated photon pair generated via spontaneous four-wave mixing is the one-dimensional Fourier transform of the product of the third-order nonlinear susceptibility and the longitudinal detuning function.
For biphoton waveform dominated by the third-order nonlinear susceptibility, damped Rabi oscillations appear in two-photon coincidence counts \cite{wen.PRA.76.013825.2007,Harris.PRL.97.113602.2006}; in contrast, biphoton waveform dominated by the longitudinal detuning function exhibits a group delay profile in such counts \cite{wen.pra.74.023808.2006,wen.pra.74.023809.2006,du.pra.93.033815.2016,du.optica.2.2014}.
Returning to Eq. (\ref{eq8}), the two-dimensional Fourier transform implies that the $\mathfrak{B}(\tau_{12},\tau_{13})$ can in principle be governed by the following four cases:
(1) $\mathfrak{B}(\tau_{12},\tau_{13})$ is dominated by $\chi^{(5)}$ in both the $\tau_{12}$ and $\tau_{13}$ axes; (2) $\mathfrak{B}(\tau_{12},\tau_{13})$ is dominated by $\Phi(\Delta kL)$ in both the $\tau_{12}$ and $\tau_{13}$ axes; (3) $\mathfrak{B}(\tau_{12},\tau_{13})$ is dominated by $\chi^{(5)}$ in the $\tau_{12}$ axis and by $\Phi(\Delta kL)$ in the $\tau_{13}$ axis; (4) $\mathfrak{B}(\tau_{12},\tau_{13})$ is dominated by $\Phi(\Delta kL)$ in the $\tau_{12}$ axis and by $\chi^{(5)}$ in the $\tau_{13}$ axis.

Analogously, if one photon is traced out, the conditional two-photon coincidence rate of the remaining two photons can be expressed as \cite{wen.PRA.77.033816.2008,wen.74.023809.2006,wu.pra.112.013706.2025}:
\begin{align}\label{eq9}
R_{cc}(\tau_{12})&=|\langle0|\hat{E}^{(+)}_{s2}\hat{E}^{(+)}_{s1}|\Psi\rangle|^2 \nonumber \\
&=A_2\int d\delta_3 |\int d\delta_2\chi^{(5)}\Phi(\Delta kL)e^{-i\delta_2\tau_{12}}|^2,  \nonumber \\
R_{cc}(\tau_{13})&=|\langle0|\hat{E}^{(+)}_{s3}\hat{E}^{(+)}_{s1}|\Psi\rangle|^2 \nonumber \\
&=A_3\int d\delta_2 |\int d\delta_3\chi^{(5)}\Phi(\Delta kL)e^{-i\delta_3\tau_{13}}|^2,
\end{align}
where $A_2$ and $A_3$ are grouped constant.

\section{Optical response of a six-level system}\label{num3}
The triphoton waveform is co-dominated by the fifth-order nonlinear susceptibility and the longitudinal detuning function.
Thus, within this section, we will investigate the linear and nonlinear optical responses of the generated signals.
On the one hand, the fifth-order nonlinear susceptibility is of crucial importance.
It not only characterizes the strength of the SSWM process, but also reveals the generation mechanism of triphotons.
On the other hand, the longitudinal detuning function determines the natural spectral width of the triphotons by regulating parameters such as group velocity.
Their competition in the $\delta_2$ and $\delta_3$ axes governs the profile of the triphoton waveform in the time domain.

The Heisenberg operator evolution equation satisfied by atomic operators $\rho_{ij}=|i\rangle\langle j|$, under the rotating-wave and dipole approximations \cite{wen.PRA.76.013825.2007}, is given by
\begin{align}\label{eq10}
\dot{\rho}_{21}&=\Gamma_{21}\rho_{21}-i\Omega^*_{p}+i(\hat{\Omega}^{(-)}_{s3}\rho_{26}-\Omega_{c1}\rho_{31}),            \nonumber \\
\dot{\rho}_{31}&=\Gamma_{31}\rho_{31}+i(\Omega^*_{p}\rho_{32}-\Omega^*_{c1}\rho_{21}+g^{(-)}_{s3}\rho_{36}-g^{(-)}_{s1}\rho_{41}),   \nonumber \\
\dot{\rho}_{41}&=\Gamma_{41}\rho_{41}+i(\Omega^*_{p}\rho_{42}-g^{(+)}_{s1}\rho_{31}+g^{(-)}_{s3}\rho_{46}-\Omega_{c2}\rho_{51}),  \nonumber \\
\dot{\rho}_{51}&=\Gamma_{51}\rho_{51}+i(\Omega^*_{p}\rho_{52}-\Omega^*_{c2}\rho_{41}-g^{(-)}_{s2}\rho_{61}+g^{(-)}_{s3}\rho_{56}),  \nonumber \\
\dot{\rho}_{61}&=\Gamma_{61}\rho_{61}+i(\Omega^*_{p}\rho_{62}-g^{(-)}_{s3}-g^{(+)}_{s2}\rho_{51}),            \nonumber \\
\dot{\rho}_{32}&=\Gamma_{32}\rho_{32}+i(\Omega_{p}\rho_{31}-g^{(-)}_{s1}\rho_{42}),  \nonumber \\
\dot{\rho}_{42}&=\Gamma_{42}\rho_{42}+i(\Omega_{p}\rho_{41}-\Omega_{c2}\rho_{52}+\Omega^*_{c1}\rho_{43}-g^{(+)}_{s1}\rho_{32}),  \nonumber \\
\dot{\rho}_{43}&=\Gamma_{43}\rho_{43}+i(\Omega_{c1}\rho_{42}-\Omega_{c2}\rho_{53}), \nonumber \\
\dot{\rho}_{52}&=\Gamma_{52}\rho_{52}+i(\Omega_{p}\rho_{51}-\Omega^*_{c2}\rho_{42}+\Omega^*_{c1}\rho_{53}-g^{(-)}_{s2}\rho_{62}), \nonumber \\
\dot{\rho}_{53}&=\Gamma_{53}\rho_{53}+i(\Omega_{c1}\rho_{52}-g^{(-)}_{s2}\rho_{63}+g^{(+)}_{s1}\rho_{54}-\Omega^*_{c2}\rho_{43}), \nonumber  \\
\dot{\rho}_{54}&=\Gamma_{54}\rho_{54}+i(g^{(-)}_{s1}\rho_{53}-\Omega^{(-)}_{s2}\rho_{64}),  \nonumber \\
\dot{\rho}_{62}&=\Gamma_{62}\rho_{62}+i(\Omega_{p}\rho_{61}-g^{(-)}_{s3}\rho_{12}+\Omega^*_{c1}\rho_{63}-g^{(+)}_{s2}\rho_{52}), \nonumber \\
\dot{\rho}_{63}&=\Gamma_{63}\rho_{63}+i(\Omega_{c1}\rho_{62}-g^{(-)}_{s3}\rho_{13}+\Omega^{(+)}_{s1}\rho_{64}-g^{(+)}_{s2}\rho_{53}), \nonumber \\
\dot{\rho}_{64}&=\Gamma_{64}\rho_{64}+i(g^{(-)}_{s1}\rho_{63}-g^{(-)}_{s3}\rho_{14}+\Omega^*_{c2}\rho_{65}-g^{(+)}_{s2}\rho_{54}), \nonumber \\
\dot{\rho}_{65}&=\Gamma_{65}\rho_{65}+i(\Omega_{c2}\rho_{64}-g^{(-)}_{s3}\rho_{15}),
\end{align}
where $\Omega_p=d_{21}E_{p0}/\hbar$, $\Omega_{c1}=d_{32}E_{c10}/\hbar$, $\Omega_{c20}=d_{54}E_{c2}/\hbar$, $g^{(+)}_{s1}=d_{34}E^{(+)}_{s1}/\hbar$, $g^{(+)}_{s2}=d_{56}E^{(+)}_{s2}/\hbar$, and $g^{(+)}_{s3}=d_{61}E^{(+)}_{s3}/\hbar$ denote the Rabi frequencies of the pump, coupling fields $E_{c1}$ and $E_{c2}$, and signals $E_{s1}$, $E_{s2}$, $E_{s3}$, respectively.
$d_{ij}=\langle i|d|j\rangle$ is the dipole matrix element.
In Eq. (\ref{eq10}), $\Gamma_{61}=i\delta_3-\gamma_{61}$, $\Gamma_{51}=i(\Delta_p+\Delta_{c1}+\delta_3+\delta_2)-\gamma_{51}$, $\Gamma_{41}=i(\Delta_p+\Delta_{c1}+\delta_3+\delta_2)-\gamma_{41}$, $\Gamma_{21}=i\Delta_p-\gamma_{21}$, $\Gamma_{31}=i(\Delta_p+\Delta_{c1})-\gamma_{31}$, $\Gamma_{32}=i\Delta_{c1}-\gamma_{32}$, $\Gamma_{42}=i(\Delta_{c1}+\delta_3+\delta_2)-\gamma_{42}$, $\Gamma_{43}=-i(\delta_3+\delta_2)-\gamma_{43}$, $\Gamma_{52}=i(\Delta_{c1}+\delta_3+\delta_2)-\gamma_{52}$, $\Gamma_{53}=-i(\delta_3+\delta_2)-\gamma_{53}$, $\Gamma_{54}=i\Delta_{c2}-\gamma_{54}$, $\Gamma_{62}=i(-\Delta_p+\delta_3)-\gamma_{62}$, $\Gamma_{63}=-i\delta_3-\gamma_{63}$, $\Gamma_{64}=-i\delta_2-\gamma_{64}$, and $\Gamma_{65}=i\delta_2-\gamma_{65}$.
$\gamma_{ij}$ ($i\neq j$) is dephasing rate between $|i\rangle$ and $|j\rangle$.

In the solution process, we adopt the weak pumping and large frequency detuning approximation, denoted as $|\Omega_{p}|^2\ll\{\Delta^2_{p},\gamma_{21}\gamma_{31}\}$.
Mathematically, the use of this approximation can greatly simplify the solution steps.
Physically, the large frequency detuning far from resonance ensures that most of the atomic population remains in the ground state, which may suppress quantum atomic noise.
It is worth noting that we do not introduce quantum Langevin noise operators into the evolution equations, as their inclusion would exclusively give rise to uncorrelated triphotons in the triphoton limit.
Specifically, upon introducing the Langevin noise operator into Eq. (\ref{eq10}), we are able to derive the corresponding Langevin noise coefficients.
Subsequently, within the Heisenberg picture and under the slowly varying amplitude approximation, we establish the coupled Heisenberg-Langevin equations for the operators ($\hat{a}_{k_{s1}}, \hat{a}_{k_{s2}}, \hat{a}_{k_{s3}}$).
Via numerical solution of these nonlinear coupled equations, we obtain the operator expressions corresponding to the output surface; from these expressions, we further derive the triple-coincidence count in the Heisenberg representation.
Note that in the coupled Heisenberg-Langevin equations, the Langevin force operators and mode operators $\hat{a}_{k_{sj}}$ are mutually decoupled--analogous to the scenario in biphoton systems \cite{du.pra.93.033815.2016,scully.pra.76.043822.2007,du.pra.107.053703.2023,du.pra.107.053703.2023r}.
As a result, these operators only generate uncorrelated triphotons, which correspond to the uniformly distributed background noise component in triple-coincidence counts and thus fall outside the scope of the present study.

By solving these equations under steady-state conditions, we derive the expressions for three quantities of interest, denoted as
\begin{align}\label{eq11}
\rho_{61}&=\frac{i\Omega_p\Omega_{c1}\Omega_{c2}g^{(-)}_{s1}g^{(-)}_{s2}}{D(\delta_2,\delta_3)(\Gamma_{21}\Gamma_{31}+|\Omega_{c1}|^2)}+\frac{ig^{(+)}_{s3}}{\Gamma_{61}}, \nonumber \\
\rho_{56}&=\frac{i\Gamma_{61}\Omega_p\Omega_{c1}\Omega_{c2}g^{(-)}_{s1}g^{(-)}_{s3}}{\Gamma^*_{61}D(\delta_2,\delta_3)(\Gamma_{21}\Gamma_{31}+|\Omega_{c1}|^2)}+o(g^{(+)}_{s3}g^{(-)}_{s3})g^{(+)}_{s2}, \nonumber \\
\rho_{34}&=\frac{i\Omega_p\Omega_{c1}\Omega_{c2}g^{(-)}_{s2}g^{(-)}_{s3}}{D(\delta_2,\delta_3)(\Gamma_{21}\Gamma_{31}+|\Omega_{c1}|^2)}    \nonumber  \\
&+\frac{i|\Omega_p|^2|\Omega_{c1}|^2\Gamma_{51}g^{(+)}_{s1}}{(\Gamma_{41}\Gamma_{51}+|\Omega_{c2}|^2)(\Gamma_{21}\Gamma_{31}+|\Omega_{c1}|^2)^2}, \nonumber \\
&D(\delta_2,\delta_3)= \Gamma_{61}(\Gamma_{41}\Gamma_{51}+|\Omega_{c2}|^2),
\end{align}
where $o(g^{(+)}_{s3}g^{(-)}_{s3})$ represents an infinitesimal quantity dependent on $g^{(+)}_{s3}g^{(-)}_{s3}$.

The slowly varying amplitudes of polarization operators for a medium are given by $P_{s1}=N\rho_{34}d_{34}$, $P_{s2}=N\rho_{56}d_{56}$ and $P_{s3}=N\rho_{16}d_{16}$.
It is known that the relationship between the polarization and the susceptibility is $P=\varepsilon_0\chi E+\varepsilon_0\chi^{(5)}EEEEE$, where $\chi$ is the linear susceptibility.
For example, for the signal $E_{s1}$, $P_{s1}=\varepsilon_0\chi_{s1}E^{(+)}_{s1}+\varepsilon_0\chi^{(5)}E_{p0}E_{c10}E_{c20}E^{(-)}_{s2}E^{(-)}_{s3}$.
Thus, one can obtain
\begin{align}\label{eq12}
\chi^{(5)}(\delta_2,\delta_3)&=\frac{iNd_{21}d_{32}d_{34}d_{54}d_{56}d_{61}}{\varepsilon_0\hbar^5D(\delta_2,\delta_3)(\Gamma_{21}\Gamma_{31}-|\Omega_{c1}|^2)}, \nonumber \\
\chi_{s1}(\delta_2,\delta_3)&=\frac{i N|d_{43}|^2|\Omega_p|^2|\Omega_{c1}|^2\Gamma_{51}}{\varepsilon_0\hbar(\Gamma_{41}\Gamma_{51}-\left|\Omega_{c2}\right|^2)(\Gamma_{21}\Gamma_{31}-\left|\Omega_{c1}\right|^2)^2}  \nonumber \\
\chi_{s2}(\delta_2)&\approx 0,~~~~ \chi_{s3}(\delta_3)=\frac{-iN|d_{16}|^2}{\varepsilon_0\hbar\Gamma^*_{61}}.
\end{align}
In our six-level system, there is at most one field between any two energy levels, which ensures that the linear and nonlinear optical responses of the generated signal derived from the density matrix model are consistent with those derived from the probability amplitude model \cite{du.josab.25.12.2008,wen.pra.74.023808.2006}.
Interested readers may attempt to use the latter model to perform validation.

\begin{figure}[t]
\centering
  \includegraphics[width=7.5cm]{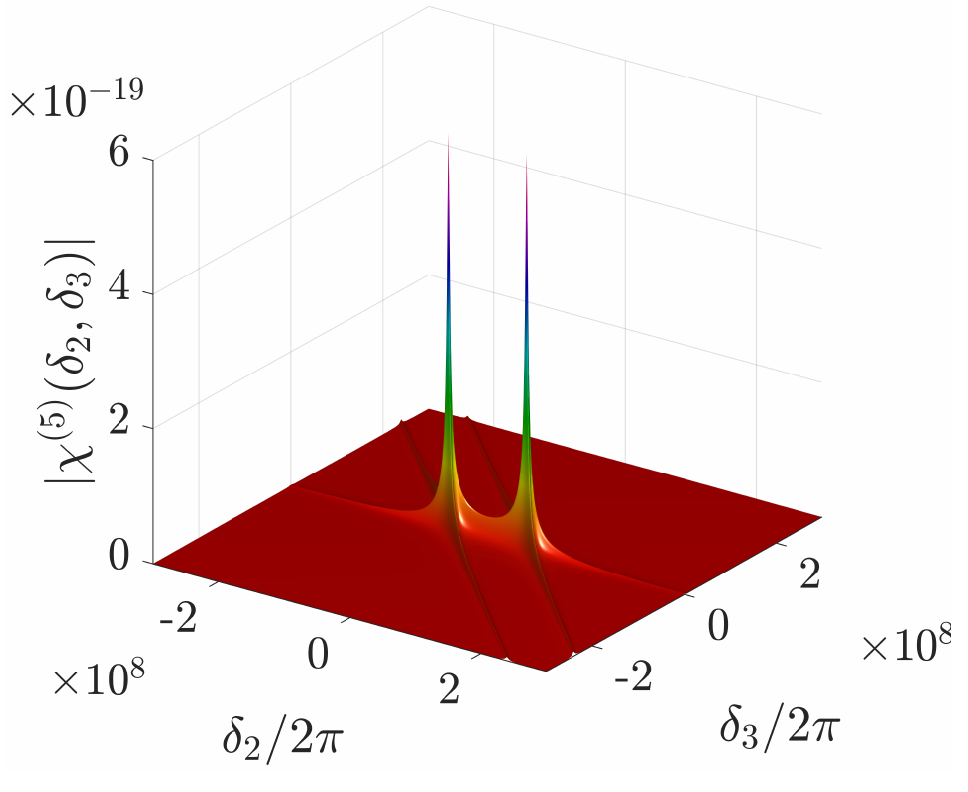}  
  \caption{The fifth-order nonlinear susceptibility $|\chi^{5}(\delta_2,\delta_3)|$ for $N=8.36\times10^{14}$, $\gamma_{21}=2\pi\times3$MHz, $\Omega_{c1}=20\gamma_{21}$, $\Omega_{c2}=20\gamma_{21}$, $\gamma_{41}=\gamma_{61}=\gamma_{21}$, $\gamma_{51}=0.2\gamma_{21}$, $\Delta_p=-2\pi\times1000$MHz, $\Delta_{c1}=-\Delta_p$, and $\Delta_{c2}=0$.}
  \label{fig2}
\end{figure}
We focus on the optical response of atomic dipole oscillations at the frequency $\varpi_{s1}+\delta_2+\delta_3$, a quantity critical for characterizing both the strength of the nonlinear parametric process and the generation mechanism of triplets.
The term $(\Gamma_{21}\Gamma_{31}-|\Omega_{c1}|^2)$ in $\chi^{(5)}(\delta_2,\delta_3)$ reveals that the two-photon resonance condition $\Delta_p+\Delta_{c1}=0$ can effectively enhance the strength of the fifth-order nonlinear parametric process.
We therefore set $\Delta_p=-\Delta_{c1}$ in the subsequent discussion.

Figure \ref{fig2} shows the fifth-order nonlinear susceptibility $|\chi^{5}(\delta_2,\delta_3)|$ for $N=8.36\times10^{14}$, $\gamma_{21}=2\pi\times3$MHz, $\Omega_{c1}=20\gamma_{21}$, $\Omega_{c2}=20\gamma_{21}$, $\gamma_{41}=\gamma_{61}=\gamma_{21}$, $\gamma_{51}=0.2\gamma_{21}$, $\Delta_p=-2\pi\times1$GHz, $\Delta_{c1}=-\Delta_p$, and $\Delta_{c2}=0$.
A distinct double-resonance structure is evident, which can be readily understood from the underlying physics.
For the denominator of $\chi^{(5)}(\delta_2,\delta_3)$, setting its real part to $\mathrm{Re}[D(\delta_2,\delta_3)]=0$ is equivalent to $\Gamma_{61}=0$ or $\Gamma_{41}\Gamma_{51}+|\Omega_{c2}|^2=0$.
Remind that there are only two independent variables in $D(\delta_2,\delta_3)$, $\delta_2$ and $\delta_3$, and the rest are constants.
The conditions $\Gamma_{61}=0$ gives rise to a single resonance peak at $\delta_3=0$ along the $\delta_3$ axis with a linewidth of $\gamma_{61}$.
Substituting this result into $\Gamma_{41}\Gamma_{51}+|\Omega_{c2}|^2=0$, we find two distinct resonances along the $\delta_2$ axis at $\delta_2=-\Omega_e/2$ and $\delta_2=\Omega_e/2$ with the linewidth of $\gamma_e$, where $\Omega_e=\sqrt{4|\Omega_{c2}|^2-(\gamma_{41}-\gamma_{51})^2}$ is the effective coupling Rabi frequency and $\gamma_e=(\gamma_{41}+\gamma_{51})/2$ is the effective dephasing rate.
This spectral feature demonstrates that our six-level atomic system supports two sets of SSWM.
For the first set, the real frequency of the $E_{s1}$, $E_{s2}$, and $E_{s3}$ photons are $\varpi_{s1}+\Omega_{e}/2$, $\varpi_{s2}-\Omega_{e}/2$, and $\varpi_{s3}$, respectively.
For the second set, these frequencies are $\varpi_{s1}-\Omega_{e}/2$, $\varpi_{s2}+\Omega_{e}/2$, and $\varpi_{s3}$, respectively.
As expected, both sets of SSWM satisfy the energy conservation condition.
The presence of these two SSWM sets implies that the triplet photons are also entangled in the frequency domain.
Notably, the energy conservation condition $\delta_1+\delta_2+\delta_3=0$ implies that $|\chi^{(5)}(\delta_2,\delta_3)|$ must exhibit a centrosymmetric structure at certain points.
Note that $\Omega_{c2}=\Omega_{c1}=20\gamma_{21}$ in Fig. \ref{fig2} is chosen solely for theoretical visualization.
In realistic cold-atom experiments, weaker Rabi frequencies are typically used to avoid perturbing the atomic spatial distribution.
The qualitative nonlinear response and triphoton generation mechanism remain unchanged for experimentally feasible, lower Rabi frequencies.

The propagation constants of the three generated signals in the atomic medium are given by $k_{s1}=(\varpi_{s1}+\delta_2+\delta_3)/\nu_{s1}$, $k_{s2}=(\varpi_{s2}-\delta_2)/\nu_{s2}$ and $k_{s3}=(\varpi_{s3}-\delta_3)/\nu_{s3}$, where $\nu_{si}$ denotes the group velocity of signal $E_{si}$ in the medium, as defined by
\begin{align}
\nu_{si}=c/[n_{si}+\omega_{si}(dn_{si}/d\delta_i)].
\end{align}
$n_{si}=\sqrt{1+\mathrm{Re}[\chi_{si}]}$ represents the refractive index experienced by the corresponding generated signal.
For signal $E_{s2}$, its group velocity equals the speed of light $c$ in vacuum.
Under the configuration of weak pump and large frequency detuning $|\Omega_{p}|^2\ll\{\Delta^2_{p},\gamma_{21}\gamma_{31}\}$, the group velocity of signal $E_{s1}$ is also approximately $c$.
Considering that $\mathrm{Re}[\chi_{s3}]\ll1$, therefore, based on the first-order Taylor expansion, we can obtain the group velocity of $E_{s3}$ as
\begin{align}
\nu_{s3}=\frac{c}{1+\frac{\varpi_{s3}N|d_{16}|^2}{2\varepsilon_0\hbar\gamma_{61}^2}}\approx \frac{2c\varepsilon_0\hbar\gamma_{61}^2}{\varpi_{s3}N|d_{16}|^2}.
\end{align}
In the propagation geometry shown in Fig. \ref{fig1}(b), the wavenumber mismatch can be approximated as $\Delta k\approx 2(\omega_{26}-\Delta_p)/c+\delta_3(1/c+1/\nu_{s3})$.

As shown in Eq. (\ref{eq9}), the triphoton waveform corresponds to the two-dimensional Fourier transform of the product of $\chi^{(5)}$ and $\Phi(\Delta kL)$.
From $\chi^{(5)}$, we extract the frequency parameter $\gamma_{61}$ along the $\delta_3$ axis, as well as $\Omega_e$ and $\gamma_e$ along the $\delta_2$ axis.
In the longitudinal detuning function $\Phi(\Delta kL)$, only the $c\nu_{s3}/[L(c+\nu_{s3})]$ term appears along the $\delta_3$ axis.
From Fourier transform theory, a narrower wavepacket in the frequency domain corresponds to a broader wavepacket in the time domain.
This implies that the contour of the triphoton waveform along the $\tau_{12}$ axis is dominated by $\chi^{(5)}$, whereas the contour along the $\tau_{13}$ axis is governed by the competition between $2\gamma_{61}$ and $c\nu_{s3}/[L(c+\nu_{s3})]$.

Here, $\nu_{s3}$ can only be dynamically adjusted by the atomic density $N$.
However, the imaginary part of the linear susceptibility, $\mathrm{Im}[\chi_{s3}(\delta_3=0)]=N|d_{16}|^2/(\varepsilon_0\hbar\gamma_{61})$, indicates that a larger atomic density $N$ results in stronger absorption of the $E_{s3}$ signal by the atomic medium, an effect detrimental to triphoton generation.
Thus, we assume the atomic density $N$ is sufficiently small to ensure the triphoton waveform is determined by $\chi^{(5)}$ along both the $\tau_{12}$ and $\tau_{13}$ axes, that is, the damped Rabi oscillation regime.

\begin{figure*}[htpb]
\centering
  \includegraphics[width=17.5cm]{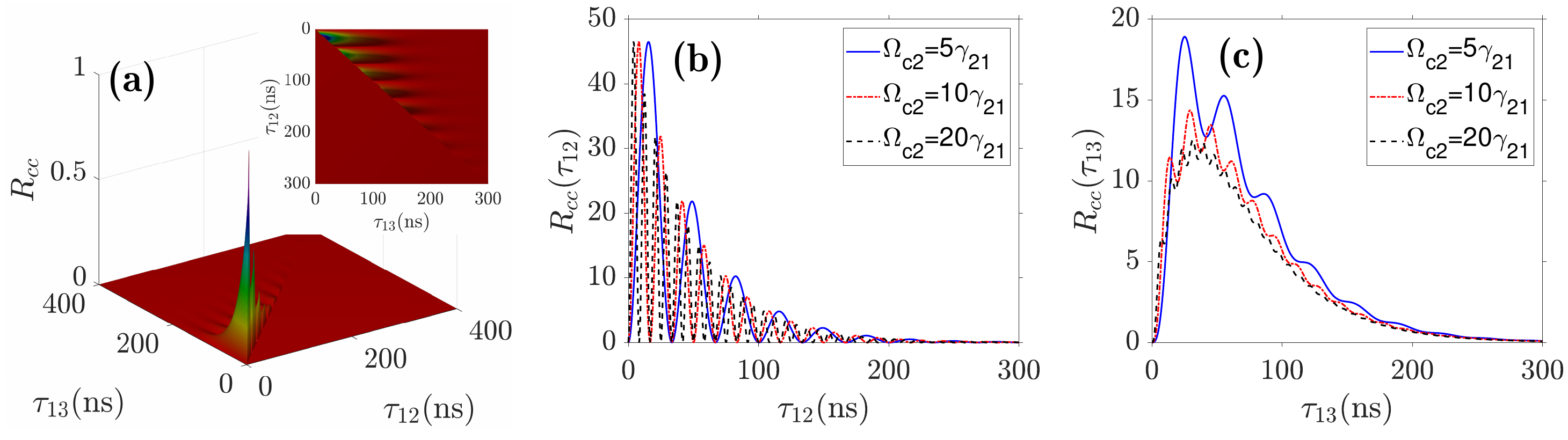}  
  \caption{(a) Normalized threefold coincidence counts in the damped Rabi regime, with parameters identical to those in Fig. \ref{fig2} except for $\Omega_{c1}=\Omega_{c2}=5\gamma_{21}$. In this regime, conditional two-photon coincidence counts $R_{cc}(\tau_{12})$ (b) and $R_{cc}(\tau_{13})$ (c) for $\Omega_{c1}=\Omega_{c2}=20\gamma_{21}, 10\gamma_{21}, 5\gamma_{21}$, where the remaining parameters are the same as (a).}
  \label{fig3}
\end{figure*}
\section{Damped Rabi oscillation regime}\label{num4}
In the damped Rabi oscillation regime, the optical properties of $\mathfrak{B}(\tau_{12},\tau_{13})$ are determined by $\chi^{(5)}$, and $\Phi(\Delta kL)$ can be approximated as unity, as explained in the Appendix.
The triphoton waveform in Eq. (\ref{eq8}) now corresponds to the two-dimensional Fourier transform of $\chi^{(5)}$,
\begin{align}\label{eq13}
&\mathfrak{B}(\tau_{12},\tau_{13})
= B_3\int\int d\delta_2d\delta_3\frac{1}{(\delta_3-i\gamma_{61})}         \nonumber \\
&\frac{e^{-i(\delta_2\tau_{12}+\delta_3\tau_{13})}}{[4(\delta_3+\delta_2-i\gamma_{51})(\delta_3+\delta_2-i\gamma_{41})-|\Omega_{c2}|^2]},
\end{align}
where $B_3$ is a grouped constant.
By calculating the residues in Eq. (\ref{eq13}), we obtain
\begin{align}\label{eq14}
\mathfrak{B}(&\tau_{12},\tau_{13})=B_4(e^{i\Omega_e\tau_{21}/2-\gamma_{e}\tau_{21}-\gamma_{61}(\tau_{13}-\tau_{12})}                                           \nonumber \\
&-e^{-i\Omega_e\tau_{21}/2-\gamma_{e}\tau_{21}-\gamma_{61}(\tau_{13}-\tau_{12})})\Theta(\tau_{13}-\tau_{12}),
\end{align}
where $B_4$ has absorbed all constant and slowly varying terms.
$\Theta(x)$ is the heaviside step function, which is defined as $\Theta(x)=1$ for $x>0$, and $\Theta(x)=0$ for $x<0$.
This clarifies the timing of triplet photon generation.
Specifically, the arrival time of the stop photon $E_{s2}$ is consistently earlier than that of $E_{s3}$.
This is also intuitive, as the transition from level $|5\rangle$ to level $|1\rangle$ can only emit $E_{s2}$ photons first, with $E_{s3}$ photons emitted subsequently.
Physically, Eq. (\ref{eq14}) describes that the time-frequency entangled nature of the triphoton waveform prohibits its factorization into products of single-photon temporal functions $f(t_1)f(t_2)f(t_3)$ or a combination of single- and two-photon correlation functions $f(t_i)f(t_j,t_k)$.
Here $f(t_i)$ is a single-photon function dependent exclusively on $t_i$, while $f(t_i,t_j)$ denotes a two-photon function characterizing the temporal correlation between $t_i$ and
$t_j$, distinct from separable single-photon dynamics.
The first term on the right-hand side of Eq. (\ref{eq14}) represents the triphoton waveform for $E_{s1}$, $E_{s2}$, and $E_{s3}$ photons with peak frequencies at $\varpi_{s1}-\Omega_e/2$, $\varpi_{s2}+\Omega_e/2$, and $\varpi_{s3}$, respectively, while the second term corresponds to the triphoton waveform where $E_{s1}$, $E_{s2}$, and $E_{s3}$ photons have peak frequencies at $\varpi_{s1}+\Omega_e/2$, $\varpi_{s2}-\Omega_e/2$, and $\varpi_{s3}$, respectively.
These two triphoton waveforms exhibit quantum destructive interference.

The threefold coincidence count rate is given by
\begin{align}\label{eq15}
R_{cc}(\tau_{12},\tau_{13})&=B_5e^{-2\gamma_{e}\tau_{12}-2\gamma_{61}(\tau_{13}-\tau_{12})}[1-\cos(\Omega_{e}\tau_{12})]   \nonumber \\
&\times\Theta^2(\tau_{13}-\tau_{12}),
\end{align}
with $B_5$ being a grouped constant.

Figure \ref{fig3}(a) shows the normalized threefold coincidence counts $R_{cc}(\tau_{12},\tau_{13})$ in the damped Rabi oscillation regime, where the parameters are chosen identically to those in Fig. \ref{fig2}, except that $\Omega_{c1}=\Omega_{c2}=5\gamma_{21}$.
The double resonance in the $\delta_2$ axis causes $R_{cc}(\tau_{12},\tau_{13})$ to exhibit damped Rabi oscillations along the $\tau_{12}$ axis.
In the absence of timing constraints on the generation of $E_{s2}$ and $E_{s3}$, a single resonance peak in the $\delta_3$ axis would lead to an exponential decay of $R_{cc}(\tau_{12},\tau_{13})$ along the $\tau_{13}$ axis with a decay rate of $2\gamma_{61}$.
Surprisingly, the existence of generation timing constraints not only significantly extends the temporal coherence between photons $E_{s1}$ and $E_{s3}$, but also restricts the damped Rabi oscillations between $E_{s1}$ and $E_{s2}$ to the region where $\tau_{13}\geq\tau_{12}$, as shown in Fig. \ref{fig3}(a).
Overall, the combined effects of the two sets of SSWM and the generation timing constraints cause $R_{cc}(\tau_{12},\tau_{13})$ to exhibit asymmetric damped Rabi oscillations with oscillation frequency $\Omega_e$.

The conditional two-photon coincidence counting rates are
\begin{align}\label{eq16}
&R_{cc}(\tau_{12})=\frac{B_5e^{-2\gamma_{e}\tau_{12}}[1-\cos(\Omega_{e}\tau_{12})]}{2\gamma_{61}},        \nonumber \\
&R_{cc}(\tau_{13})=B_5\frac{(e^{-2\gamma_{e}\tau_{13}}-e^{-2\gamma_{61}\tau_{13}})}{2\gamma_{6e}}\Theta(\tau_{13})     \nonumber \\
&-B_5\frac{[-2e^{-2\gamma_{61}\tau_{13}}\gamma_{6e}+e^{-2\gamma_{e}\tau_{13}}\Omega_e\sin(\Omega_e\tau_{13})]}{\Omega_e^2+4\gamma_{6e}^2}\Theta(\tau_{13})  \nonumber \\
&-B_5\frac{2e^{-2\gamma_{e}\tau_{13}}\gamma_{6e}\cos(\Omega_e\tau_{13})}{\Omega_e^2+4\gamma_{6e}^2}\Theta(\tau_{13}),    \nonumber \\
&R_{cc}(\tau_{23})=\frac{B_5\Omega_{e}^2}{2\gamma_{e}(\Omega_{e}^2+4\gamma_{e}^2)}e^{-2\gamma_{61}\tau_{23}},
\end{align}
where $\gamma_{6e}=\gamma_{61}-\gamma_{e}$.
$R_{cc}(\tau_{12})$ exhibits damped Rabi oscillations along the $\tau_{12}$ axis, consistent with the coincidence counts of correlated photon pairs generated by spontaneous four-wave mixing in a four-level system \cite{du.josab.25.12.2008,wen.pra.74.023808.2006}.
This is also readily understandable.
When we trace out the $E_{s3}$ photon, the remaining optical transition pathway matches that of the four-level system, where $E_p$ and $E_{c1}$ are treated as a single entity.
Figure \ref{fig3}(b) shows the evolution of $R_{cc}(\tau_{12})$ as a function of $\tau_{12}$ for different $\Omega_{c2}$.
The oscillation frequency is inherited from the triphoton waveform.
When $\Omega_{c2}=5\gamma_{21}$, the oscillation period is 33 ns, while for $\Omega_{c2}=10\gamma_{21}$ and $\Omega_{c2}=20\gamma_{21}$, the oscillation periods are one-half and one-quarter of that at $\Omega_{c2}=5\gamma_{21}$, respectively.
$R_{cc}(\tau_{13})$ is a superposition of multiple exponential decay functions and damped Rabi oscillations, as shown in Fig. \ref{fig3}(c), whose oscillation frequency corresponds to the $\Omega_e$.
It is worth noting that the oscillations in the profile of $R_{cc}(\tau_{13})$ gradually diminish as $\Omega_{c2}$ increases, until they evolve into a single-peaked exponential decay.
$R_{cc}(\tau_{23})$ exhibits a simple exponential decay with a decay rate of $2\gamma_{61}$.
It is important to emphasize that, in conditional two-photon coincidence counts, the absence of oscillations does not invalidate the triphoton character of the process.
Even when no oscillations appear, as in $R_{cc}(\tau_{23})$, the underlying process remains a genuine triphoton generation mechanism.
Oscillations are only a characteristic signature of temporal correlations under specific conditions, not a necessary requirement for the process to be identified as triphoton.

\begin{figure}[t]
\centering
  \includegraphics[width=7.5cm]{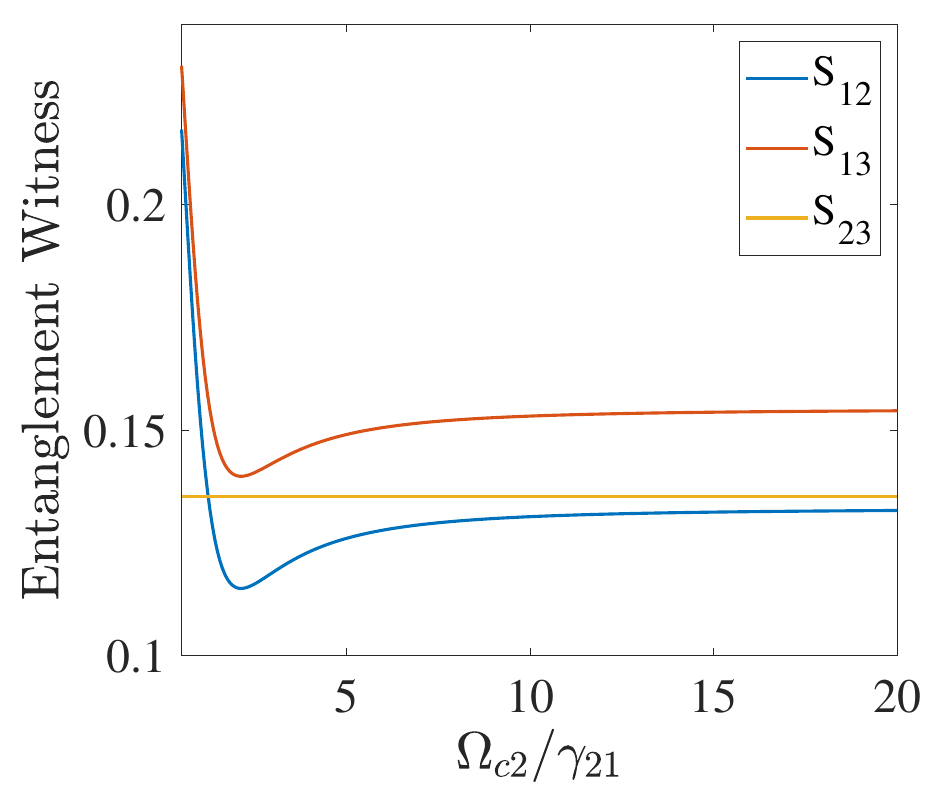}  
  \caption{The entanglement criterion $S_{ij}$ evolves with $\Omega_{c2}$, while the other parameters remain the same as in Fig. \ref{fig2}.}
  \label{fig4}
\end{figure}

Finally, we concisely verify that the generated state is an energy-time entangled $W$-class state by invoking the energy-time entanglement criterion.
For any two modes $E_{sj}$ and $E_{sk}$, if they are separable, their bipartite joint time-frequency uncertainty product satisfies the inequality \cite{mancini.prl.88.120401.2002}
\begin{align}\label{eq17}
S_{ij}=\Delta\tau_{jk}\Delta(\omega_{sj}+\omega_{sk})\geq1,
\end{align}
where
\begin{align*}
\Delta\tau_{jk}=\sqrt{\frac{\int\tau_{jk}^2R_{cc}(\tau_{jk})d\tau_{jk}-(\int\tau_{jk}R_{cc}(\tau_{jk})d\tau_{jk})^2}{\int R_{cc}(\tau_{jk})d\tau_{jk}}},
\end{align*}
and $\Delta(\omega_{sj}+\omega_{sk})$ is defined analogously in the frequency domain.
A violation of this inequality indicates the presence of bipartite entanglement.
For the triphotons depicted in Fig. \ref{fig3}(a), we extract the temporal parameters $\Delta\tau_{12}=42$ns, $\Delta\tau_{13}=49.7$ns, and $\Delta\tau_{23}=26.5$ns.
Since $\tau_{jk}$ and $\omega_{sj}+\omega_{sk}$ are conjugate variables, the corresponding frequency-domain correlation function can be derived by normalizing the two-photon time-domain correlation function presented in Eq. (\ref{eq16}) followed by Fourier transformation.
Subsequently, we extract $\Delta(\omega_{sj}+\omega_{sk})$ via Gaussian fitting of the resulting frequency-domain correlation function, with the specific fitting parameters listed as follows:
 $\Delta(\omega_{s1}+\omega_{s2})\approx 3$MHz, $\Delta(\omega_{s1}+\omega_{s3})\approx 3$MHz, and $\Delta(\omega_{s2}+\omega_{s3})\approx 5.1$MHz.
Note that three narrow-linewidth optical cavities are utilized for spectral filtering during experimental triphoton generation \cite{li2024direct,feng2025observation}--this further suppresses the frequency-sum uncertainty of the conditional biphotons derived from the triphoton state.
Substituting these values into the entanglement criterion, we find that $S_{12}\approx0.13$, $S_{13}\approx 0.15$, and $S_{23}\approx 0.14$.
As a supplementary analysis, we examine the behavior of the entanglement criterion for other values of the parameter $\Omega_e$.
Since $\Omega_e$ is proportional to $\Omega_{c2}$, Figure \ref{fig4} depicts the evolution of $S_{ij}$ as a function of $\Omega_{c2}$.
As $\Omega_{c2}$ increases, $S_{12}$ and $S_{13}$ first decrease to a minimum, then rise and eventually saturate at a constant value.
This behavior directly reflects the gradual separation of the two resonance peaks evident in Fig. \ref{fig2}.
In contrast, $S_{23}$ remains constant over the entire parameter range, as its time-domain waveform is independent of $\Omega_{c2}$.
Notably, $S_{ij}<1$ holds for all pairs, confirming bipartite entanglement in every two-mode subsystem.
This universal bipartite entanglement among all two-mode subsystems is the hallmark signature of tripartite $W$-class entanglement \cite{PhysRevA.62.062314}, which clearly distinguishes it from other tripartite entangled states such as GHZ states.
Thus, we conclusively establish that the generated triphotons correspond to energy-time entangled tripartite $W$-class states.

\section{Conclusion}\label{num6}
In summary, we have investigated the generation mechanism of energy-time entangled triphotons in a six-level cold atomic system.
The triphoton waveform is characterized by the two-dimensional Fourier transform of the fifth-order nonlinear susceptibility and longitudinal detuning function.
By solving the Heisenberg evolution equations governing the atomic operators, we derived the linear and fifth-order nonlinear susceptibilities for each generated signal.
The linear optical response reveals that $E_{s1}$ and $E_{s2}$ photons propagate at the speed of light in the atomic medium, whereas the group velocity of $E_{s3}$ is inversely proportional to the atomic density.
Increasing atomic density results in strong absorption of the $E_{s3}$ signal by the medium.
Thus, we assume the atomic density is sufficiently low such that the triphoton waveform is determined by fifth-order nonlinear susceptibility.
The fifth-order nonlinear susceptibility reveals two sets of spontaneous six-wave mixing processes in the system, with the $E_{s2}$ photon generated prior to the $E_{s3}$ photon in each set.
These two sets of spontaneous six-wave mixing processes, together with the temporal ordering of photon generation, lead to threefold coincidence counts that exhibit asymmetric damped Rabi oscillations in the two-dimensional temporal domain,
which is distinct from the symmetric damped Rabi oscillations reported in previous works \cite{kangkang.aqt.35.2020,li2024direct} where no temporal ordering of photon generation was imposed.
Finally, the two-photon temporal correlation properties between two subsystems further confirm that the triplet photons generated in the six-level system form $W$-class entangled states in the energy-time degrees of freedom.
These findings not only lay a foundation for understanding the more complex mechanisms of triphoton generation and their optical properties but also provide a practical physical pathway for the precise preparation and manipulation of $W$-class triplet photons.
\section*{Acknowledgement}
This work was supported by the National Natural Science Foundation of China (12204293,12404415), the Fundamental Research Program of Shanxi Province (202203021212387), and Shanxi Scholarship Council of China (2025-142).
\section*{Appendix}
\subsection{Treatment of transverse detuning function}
The transverse detuning function is defined in Eq. (5) as:
\begin{align}
H(\bm{\alpha}_{s1}+\bm{\alpha}_{s2}+\bm{\alpha}_{s3},\bm{\rho})=A^{-\frac{3}{2}}\int_{A}d^2\rho e^{-i(\bm{\alpha}_{s1}+\bm{\alpha}_{s2}+\bm{\alpha}_{s3})\cdot\bm{\rho}}.   \nonumber \\
\end{align}
where $A$ is the common effective transverse area of the three input classical fields, $\bm{\alpha}_{si}$ are the transverse wavevectors of the generated $E_{si}$ photons, and $\bm{\rho}$ is the transverse position vector perpendicular to the propagation axis $z$.
When the transverse area $A$ is sufficiently large and diffraction effects are negligible, the transverse integral approaches the two-dimensional Dirac delta function in the limit:
\begin{align}
\lim_{A\rightarrow\infty}A^{-1}\int_{A}d^2\rho e^{-i\bm{K}\cdot\bm{\rho}}=\delta^{(2)}(\bm{K}),
\end{align}
where $\bm{K}=\bm{\alpha}_{s1}+\bm{\alpha}_{s2}+\bm{\alpha}_{s3}$.
Substituting this limit into our definition, we obtain:
\begin{align}
H(\bm{K},\bm{\rho})=A^{-\frac{1}{2}}(A^{-1}\int_{A}d^2\rho e^{-i\bm{K}\cdot\bm{\rho}})=A^{-\frac{1}{2}}\delta^{(2)}(\bm{K}).
\end{align}
This approximation is mathematically well-defined and physically corresponds to strict transverse momentum conservation for the triphoton generation process: only when the total transverse wavevector mismatch $\bm{K}=0$ does the function $H(\bm{K},\bm{\rho})$ yield a non-zero contribution, which is exactly the property of a Dirac delta function.
The $A^{-1/2}$ prefactor is a global normalization constant that is absorbed in the overall normalization of the triphoton state, and does not affect any physical conclusions regarding the temporal correlation discussed in the manuscript.
\subsection{Validity of the unit approximation for the longitudinal detuning function} 
Here we justify the approximation $\Phi(\Delta kL)\approx1$ for the triphoton waveform dominated by $\chi^{(5)}$.
Since the triphoton waveform corresponds to the two-dimensional Fourier transform of the product of $\chi^{(5)}$ and $\Phi(\Delta kL)$, we first analyze $\Phi(\Delta kL)$.
As shown in the main text, $\Delta k\approx 2(\omega_{26}-\Delta_p)/c+\delta_3(1/c+1/\nu_{s3})$ depends only on $\delta_3$, and its Fourier transform reads
\begin{align}
\mathfrak{B}(\tau_{13})&=B_2\int d\delta_3\Phi(\Delta kL)e^{-i\delta_3\tau_{13}} \nonumber \\
&=B_6e^{-(\omega_{26}-\Delta_p)\nu_{s3}\tau_{13}/c}\Theta\left(\tau_{13}, \frac{L}{\nu_{s3}}\right)
\end{align}
where $B_6$ is a grouped constant, and the step function $\Theta\left(\tau_{13}, L/\nu_{s3}\right)$ equals unity for $\tau_{13}\in(0, L/\nu_{s3})$ and zero otherwise.

Using the parameters given in Fig. \ref{fig3}(a), we obtain $L/\nu_{s3}\approx3$ns, which is much shorter than the characteristic time set by $\chi^{(5)}$: $1/(2\gamma_{61})\approx 26.5 \mathrm{ns}$.
In the time domain, this implies that the triphoton amplitude governed by $\chi^{(5)}$ fully envelopes the contribution from $\Phi(\Delta kL)$.
For theoretical simplicity, we therefore approximate $\Phi(\Delta kL)\approx1$ in our analysis.
It should be noted that in the above analysis, we have not accounted for factors such as the generation timing of the triplet photons, but have focused exclusively on the Fourier transform itself.

%

\end{document}